\documentclass{article}

\usepackage{arxiv}

\usepackage[utf8]{inputenc} 
\usepackage[T1]{fontenc}    
\usepackage{hyperref}       
\usepackage{url}            
\usepackage{booktabs}       
\usepackage{amsfonts}       
\usepackage{nicefrac}       
\usepackage{microtype}      
\usepackage{lipsum}		
\usepackage{graphicx}
\usepackage{natbib}
\usepackage{doi}
\usepackage{amsmath}

\title{Implementation of Lenia as a Reaction-Diffusion System}

\author{
    Hiroki Kojima\\
	Graduate School of Arts and Sciences, The University of Tokyo\\
    Alternative Machine Inc. \\
	\texttt{kojima@sacral.c.u-tokyo.ac.jp}\\
	\AND
	Takashi Ikegami \\
	Graduate School of Arts and Sciences, The University of Tokyo\\
    Alternative Machine Inc. \\
	\texttt{ikeg@sacral.c.u-tokyo.ac.jp}\\
}

\renewcommand{\shorttitle}{Implementation of Lenia as a RD System}

\hypersetup{
pdftitle={Lenia and Reaction Diffusion system},
pdfauthor={Hiroki Kojima, Takashi Ikegami},
pdfkeywords={Lenia, Reaction-diffusion system, Cellular Automata},
}

\begin{document}
\maketitle

\begin{abstract}

The relationship between reaction-diffusion (RD) systems, characterized by continuous spatiotemporal states, and cellular automata (CA), marked by discrete spatiotemporal states, remains poorly understood. This paper delves into this relationship through an examination of a recently developed CA known as Lenia. We demonstrate that asymptotic Lenia, a variant of Lenia, can be comprehensively described by differential equations, and, unlike the original Lenia, it is independent of time-step ticks. Further, we establish that this formulation is mathematically equivalent to a generalization of the kernel-based Turing model (KT model). Stemming from these insights, we establish that asymptotic Lenia can be replicated by an RD system composed solely of diffusion and spatially local reaction terms, resulting in the simulated asymptotic Lenia based on an RD system, or “RD Lenia”. However, our RD Lenia cannot be construed as a chemical system since the reaction term fails to satisfy mass-action kinetics.

\end{abstract}


\section{Introduction}

\begin{figure}
	\centering
	 \includegraphics[width=0.7\linewidth]{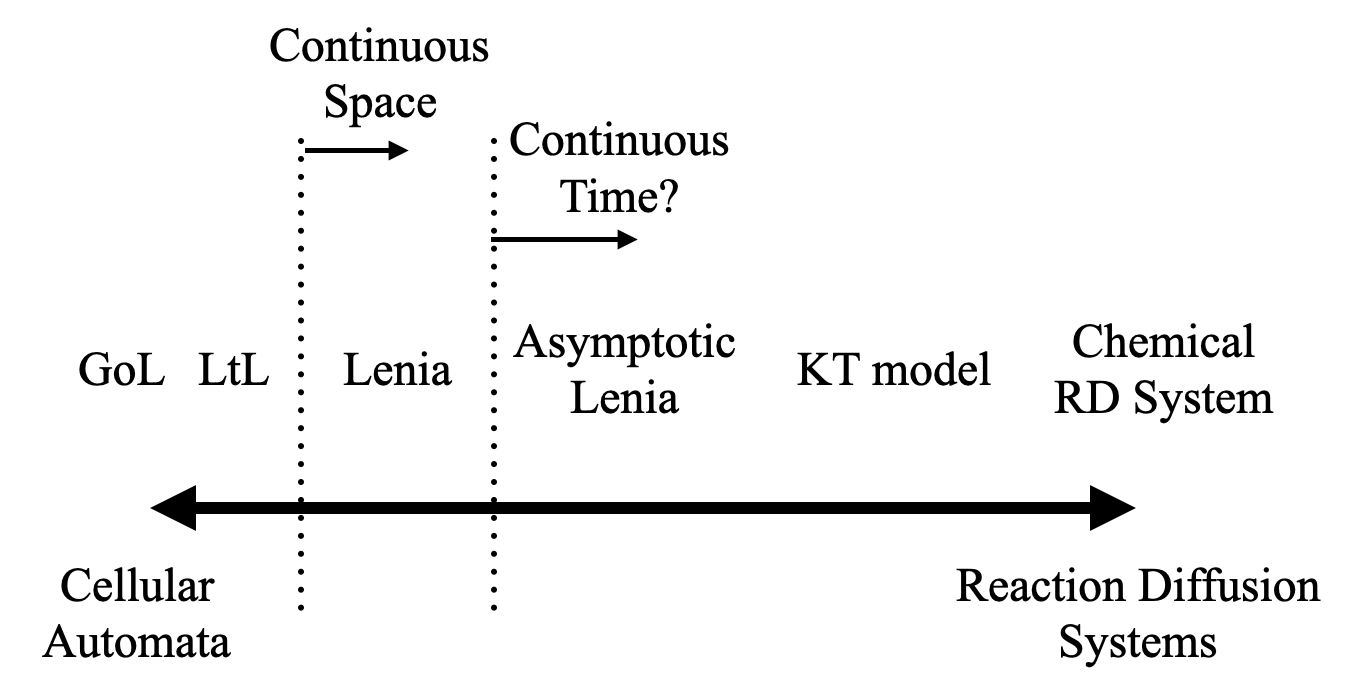}
	\caption{Overall view of two-dimensional CA and RD systems}
	\label{fig:overview}
\end{figure}

Pattern formation constitutes a key focus in artificial life research, and two primary approaches have been explored. The first approach involves reaction-diffusion (RD) systems. Classic examples include Turing patterns, \citep{turing1990chemical}, which demonstrated that chemical reaction-diffusion systems, encompassing local reaction and diffusion terms, could synthesize intricate spatial patterns. Actual chemical experiments such as the Belousov-Zhabotinsky (BZ) reaction and the iodine clock reaction, which display a range of reaction patterns including chaotic oscillation, underwent intensive study in the 1980s. The second approach involves cellular automata (CA). The most renowned two-dimensional CA is Conway's Game of Life (GoL) \citep{gardner1970GoL}, which exhibits numerous complex and unpredictable patterns. Specifically, the glider, a pattern that stably moves in one direction, is widely recognized. There exist numerous variants of GoL, for instance, Larger than Life (LtL) \citep{evans2001larger}, which displays diverse gliders by utilizing a large number of neighbors.
Lenia  \citep{chan2019lenia,chan2020lenia}, a recent variant of CA, was proposed to modify GoL in a spatially and temporally continuous manner. An intriguing aspect of Lenia is its display of extremely complex, life-like patterns that encompass both global spatial patterns, akin to those in an RD system, and stable moving patterns, like the GoL's “glider”. Based on these observations, we hypothesize that Lenia might be situated at the intersection of CA and RD systems (Fig.\ref{fig:overview}). Thus, this paper discusses the possibility of formulating Lenia as a (chemical) RD system, thereby linking CA and RD systems.

In subsequent sections, we review some related systems and outline the contributions of this study.

\subsection{RD System}

A general RD system is formulated as follows: 

\begin{equation}
\frac{\partial u}{\partial t} = D \nabla^2 u + F(u),
\end{equation}
where $u(x,t)$ is a chemical concentration at $x$ at time $t$ and $F(u)$ is a reaction term. Note that the computation of the reaction term, $F(u)$, is spatially local, and spatial information that can lead to Turing patterns is introduced only by the diffusion term $D \nabla^2 u$.

Recently, \citet{kondo2017updated} proposed that the most of the pattern from the RD systems can be emulated by non-local kernel methods (the kernel-based Turing model (KT model)), formulated as follows:

\begin{equation}
u(x,t+dt) = u(x,t) + dt([K*u]_{0}^{a_{max}} - u),
\label{eq:KTmodel}
\end{equation}
where $K(x)$ is a non-local kernel function, $*$ is a convolution operation and $[]^{a_{max}}_{0}$ is a clipping into $[0,a_{max}]$.

This implies the underlying connection between RD systems and non-local kernel models and, for one-dimensional systems, \citet{ninomiya2017reaction} proved mathematically that the dynamics generated by non-local kernel methods can be emulated by an RD system.

\subsection{Lenia and asymptotic Lenia}

Lenia was developed as a continuous version of GoL \citep{chan2019lenia,chan2020lenia} and is formulated as follows:

\begin{equation}
	u(x,t+dt) = [u(x,t) + dt G(K*u)]^{1}_{0},
 \label{eq:Lenia}
\end{equation}
where $K*u$ is a non-local kernel convolution, $[]^{1}_{0}$ is a clipping into $[0,1]$, and $G$ is a reaction term, usually using the highly localized Gaussian function $G(z) = 2\exp(-(\frac{z-m}{s})^2)-1$.

The asymptotic Lenia \citep{kawaguchi2021introducing} is a variant of Lenia and formulated as follows:

\begin{equation}
	u(x,t+dt) = u(x,t) + dt(T(K*u)-u),
 \label{eq:asymLenia}
\end{equation}
where $T$ is a reaction term and is usually constructed from the reaction term in the original Lenia $G$ as $T = \frac{(G+1)}{2}$.

Eq.\ref{eq:asymLenia} does not include the clipping procedure in Eq.\ref{eq:Lenia}. This allows us to transform Eq.\ref{eq:asymLenia} into a differential equation:

\begin{equation}
	\frac{\partial}{\partial t} u(x,t) = T(K*u)-u
 \label{eq:asymLenia_diff}
\end{equation}

\subsection{Structure of this Paper}

In this paper, we have two main findings. First, we will show that the formulation of asymptotic Lenia is mathematically equivalent to a generalized KT model \citep{kondo2017updated}. This will be shown in the first part (\S\ref{sec:2.1}), and the relationship between (asymptotic) Lenia and RD systems will be discussed.

The second main result is that the asymptotic Lenia can be implemented as an RD system. We will explain the theoretical idea and the actual implementation ("RD Lenia") in the second part (\S\ref{sec:2.2}).

Finally, we will discuss the criterion that the system can be implemented chemically and whether the RD Lenia is chemically plausible (\S\ref{sec:2.2.3}).

\section{Results}

\subsection{Lenia and RD Systems}\label{sec:2.1}

In this part, we study the properties of (asymptotic) Lenia in order to explore the possibility of translation between Lenia and RD systems. First, we show the mathematical equivalence between asymptotic Lenia and KT models, and second, we check the temporal continuity in (asymptotic) Lenia, which is a prerequisite for its formulation as an RD system in the next part.

\subsubsection{Asymptotic Lenia as a Generalization of the KT Model}\label{sec:2.1.1}

As we saw in Eq.\ref{eq:asymLenia_diff}, the asymptotic Lenia can be formulated as a differential equation, which is different from the original Lenia, which involves a clipping procedure that cannot be expressed in a differential equation. We found that this mathematical formulation of the asymptotic Lenia is equivalent to a generalized KT model. This can be proved by showing that we can recover the form of the KT model (Eq.\ref{eq:KTmodel}) from Eq.\ref{eq:asymLenia} by setting the reaction function $T$ as follows:

\begin{equation}
    T(K*u)=
    \begin{cases}
      0, &  K*u < 0  \\
      K*u, & 0 \leq K*u \leq a_{max} \\
      a_{max}, &   K*u > a_{max} 
    \end{cases}
\end{equation}

Therefore, the mathematical formulation of the asymptotic Lenia (Eq.\ref{eq:asymLenia_diff}) can be considered as a generalized formulation of the KT model. In the context of RD systems, similar kinds of generalized formulations have been discussed \citep{ninomiya2017reaction,cygan2021pattern}, and we can apply two consequences from these previous studies to the asymptotic Lenia through the equivalence we found.

First, the results of \citet{ninomiya2017reaction} showed that this kind of system can be reformulated as an RD system, which suggests that the asymptotic Lenia can also be emulated by an RD system. \citet{ninomiya2017reaction} proved this only for one-dimensional systems, so we cannot directly apply this to asymptotic Lenia. We will show that the same idea can be used for higher-dimensional systems, and construct an RD system that emulates asymptotic Lenia in the next part (\S\ref{sec:2.2}).

Second, the results of \citet{cygan2021pattern} may be related to the relationship between the shape of the reaction terms and the resulting behavior of the asymptotic Lenia. \citet{kawaguchi2021introducing} explored the region of $G$ to find nonconstant solutions by systematically changing the shape of Gaussian functions, but there is no mathematical criterion to find the appropriate reaction terms. The results of \citet{cygan2021pattern} seem to be related to this, as they provide the sufficient condition for the system described by Eq.\ref{eq:asymLenia_diff} to have a weak non-constant stationary solution. Unfortunately, we cannot simply apply this to most of the asymptotic Lenia because one of the conditions is $\lambda_k = \frac{1}{T'(0)}$, (where $\lambda_k$ is one of the eigenvalues of the kernel function $K$,) and this is hardly applicable to the common settings of the asymptotic Lenia, where $T$ is usually a highly localized Gaussian function and the gradient at the origin, $T'(0)$, is vanishing.

\subsubsection{Is the asymptotic Lenia continuous in Time?}\label{sec:2.1.2}

To emulate a system by RD systems, it should be temporally continuous. This is not the case with the original Lenia. The original Lenia was designed to be  spatio-temporally continuous, but it has been found that it yields different dynamics not only for large but also for small time steps \citep{tyrell2022step}. 
In the previous section, we showed that the asymptotic Lenia can be described by differential equations, which seems to ensure that the asymptotic Lenia is temporally continuous, unlike original Lenia. However, in the actual simulation, we have to discretize the process in the Euler method, and we cannot simply dismiss the possibility that this discretization might affect the resulting patterns. Therefore, we double-check that the asymptotic Lenia is independent of the size of the step by running the simulation changing the time step ticks of the Euler method.

\begin{figure}
\centering
\begin{tabular}{cc}
  \begin{minipage}[t]{0.45\hsize}
  \includegraphics[width=\linewidth]{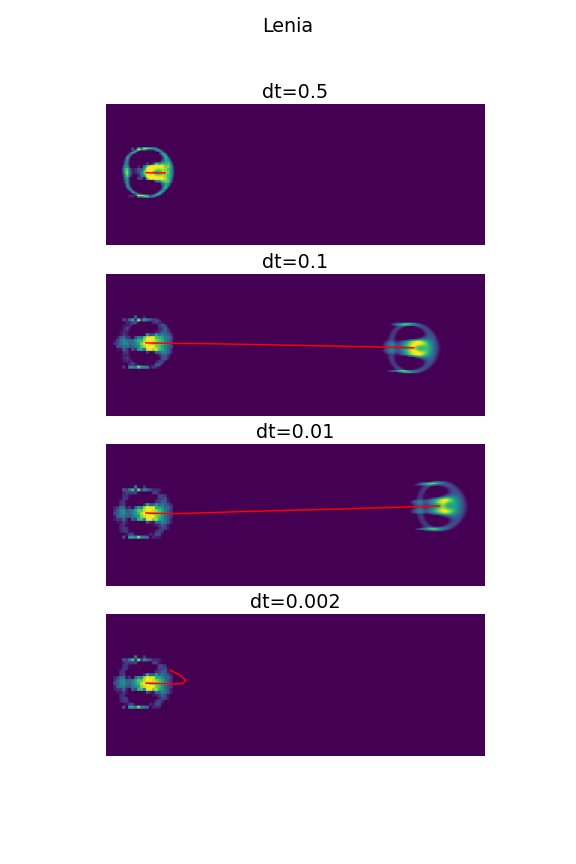}
  \end{minipage} &
 \begin{minipage}[t]{0.45\hsize}
  \includegraphics[width=\linewidth]{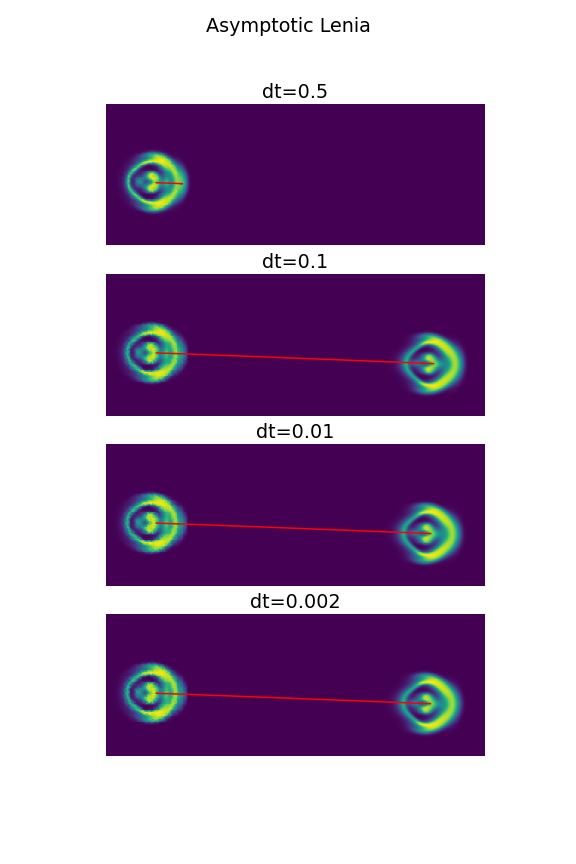}
  \end{minipage} 
\end{tabular}
\caption{Dependence on time step in Lenia (Left) and asymptotic Lenia (Right) with varying time step sizes (dt = 0.5, 0.1, 0.01, 0.002). The initial and final patterns are superimposed. The motion trajectory, represented by the red line, is calculated as the center of mass of the pattern. In instances where no final glider pattern is visible in the figure, it indicates that the glider has decayed at a certain point.}
\label{fig:timestep}
\end{figure}

First, we replicated the effect of time step size in the original Lenia (Fig.\ref{fig:timestep}, Left). Here, we used the “orbium” pattern \citep{chan2019lenia}. We observed that the pattern disappeared both when the size of the time step was large ($dt=0.5$) and when it was small ($dt=0.002$). We then checked the effect of the time step size in the asymptotic Lenia. In this case, if the size of the time step was large, the pattern disappeared, but, otherwise, the pattern was maintained even with $dt=0.002$ (Fig.\ref{fig:timestep}, Right).
Therefore, we confirmed that the asymptotic Lenia behaves continuously over time, which is different from the original Lenia.

\subsubsection{The Relationship between the Time Step Dependence in Lenia and the Clipping Procedure}\label{sec:2.1.3}

The size and time step dependence was only observed in Lenia and not in the asymptotic Lenia. The main difference in the formulation between the original Lenia and the asymptotic Lenia is the presence of the clipping procedure $[ ]_{0}^{1}$, which seems to suggest that this clipping procedure is responsible for the size dependence.

We hypothesized that the clipping procedure at certain time steps stabilizes pattern formation in Lenia, and we further hypothesized that if we apply clipping on a certain time scale, the pattern should be preserved even if the time step of the update is much faster or even continuous.  To test this, we ran Lenia under different conditions. First, the “no clipping condition” was Lenia with only the lower clipping, $[ ]_{0}^{\infty}$. Second, the “occasional clipping condition” was basically the same as the “no clipping condition”, but occasionally applied the upper clipping, $[ ]_{0}^{1}$. Our hypothesis predicts that no clipping will lead to no pattern formation, and occasional clipping will rescue it.

\begin{figure}
\centering
\includegraphics[width=0.45\linewidth]{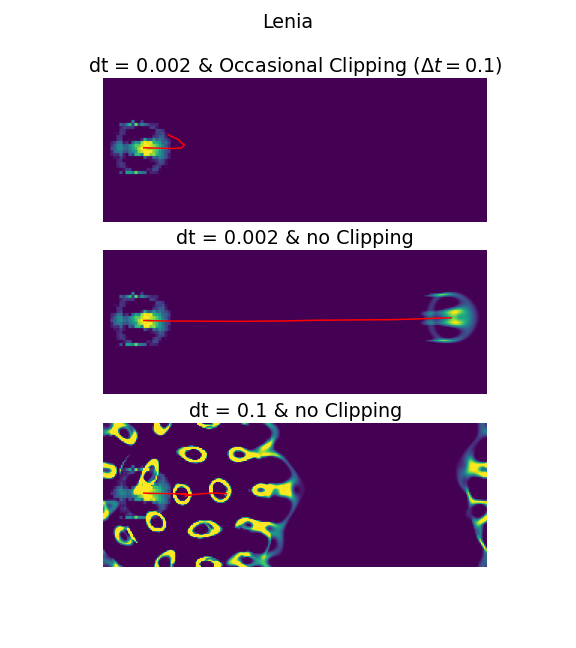}
\caption{Effect of clipping on Lenia. Top: The result of occasional clipping ($\Delta t = 0.1$), updated at $dt = 0.002$, Middle: The result of the no-clipping condition, updated at $dt = 0.002$, Bottom: The result from the no-clipping condition, updating at $dt = 0.1$.}
\label{fig:LeniaClipping}
\end{figure}

The simulation results are shown in Fig.\ref{fig:LeniaClipping}. We found that, contrary to our expectation, this occasional clipping did not support pattern maintenance (Fig.\ref{fig:LeniaClipping}, Top). Rather, when we completely omitted the upper clipping procedure, the pattern unexpectedly persisted at $dt = 0.002$, implying that the clipping procedure inhibits pattern maintenance at small time step sizes (Fig.\ref{fig:LeniaClipping}, Middle). Note that this is only valid for the case of small time steps. For example, if $dt = 0.1$ with no upper clipping, the local pattern will collapse (Fig.\ref{fig:LeniaClipping}, Bottom).

In summary, in the small time step regions, the pattern of Lenia is preserved when we omit the upper clipping, which means that when we simulate this region, Lenia behaves in a temporally continuous manner. However, in the middle time step region ($dt = 0.1$), clipping is necessary to maintain the pattern. The behavior in this region cannot be reproduced by $dt = 0.002$ with occasional clipping ($\Delta t = 0.1$). This means that the behavior of Lenia in this region cannot be simulated in a temporally continuous manner, even if we add additional clipping procedures to the system.

\subsection{Implementing the asymptotic Lenia using RD Systems (RD Lenia)}\label{sec:2.2}

In the previous part (\S\ref{sec:2.1}), we studied the properties of the asymptotic Lenia and found the relation to the model of Turing patterns (KT model) and also confirmed that the dynamics is temporally continuous. Now we investigate whether the asymptotic Lenia can be implemented by (chemical) RD systems. There are two main components: the non-local kernel and the reaction term. First, we show how the asymptotic Lenia can be attributed to an RD system that does not contain a non-local kernel, based on the idea of \citet{ninomiya2017reaction}, and second, we discuss the difficulty of interpreting the reaction term as a chemical reaction.

\subsubsection{Approximation of Non-local Kernel by RD Systems}\label{sec:2.2.1}

To implement a non-local kernel with the RD system, we follow the methods of \citet{ninomiya2017reaction}, which showed that a generalized one-dimensional RD system with non-local kernel convolution, 

\begin{equation}
\frac{\partial u}{\partial t} = D \frac{\partial^2 u}{\partial x^2} + F(u,K*u),
\label{eq:1dNinomiya}
\end{equation}
can be emulated by an RD system without a non-local kernel.

To this end, they introduced auxiliary variables $\{v_j\}$ and showed that the following RD system could emulate the system using Eq.\ref{eq:1dNinomiya}:

\begin{equation}
    \begin{cases}
      \dfrac{\partial u}{\partial t} = D\dfrac{\partial^2 u}{\partial x^2} + F(u,\alpha_0+\sum_{j=1}^{M} \alpha_j v_j)  \\
      \dfrac{\partial v_j}{\partial t} = \dfrac{1}{\epsilon} (D_j\dfrac{\partial^2 v_j}{\partial x^2} +\mu u - v_j),
    \end{cases}
\end{equation}
where $M$ is the number of auxiliary variables $\{v_j\} (j = 1,..,M)$, $D_j$ is the diffusion constant for the chemical $v_j$ and $\{\alpha_j\}, \mu, \epsilon$ are constants ($\mu , \epsilon >0 , |\epsilon| \ll 1$). In the original setting, $D_j = \frac{1}{j^2}$.

The key feature of this system is that the term $F(u,K*u)$ in Eq.\ref{eq:1dNinomiya} is substituted by $F(u,\alpha_0 + \sum_{j=1}^M \alpha_j v_j) $, which means that the non-local kernel convolution $K*u$ is approximated by the superposition of auxiliary variables $\{v_j\}$ as $\alpha_0 + \sum_{j=1}^M \alpha_j v_j$. \citet{ninomiya2017reaction} showed that any symmetric kernel can be approximated by this method. Furthermore, $\epsilon$ indicates how fast the term $\alpha_0 + \sum_{j=1}^M \alpha_j v_j$ should converge to the emulated non-local kernel, so $\epsilon$ should be much smaller compared to the time scale of the dynamics of Eq.\ref{eq:1dNinomiya}.

In the original paper, this was shown only in one-dimensional RD systems. Here, we applied the same idea to a two-dimensional generalized RD system,

\begin{equation}
\frac{\partial u}{\partial t} = D \nabla^2 u + F(u,K*u),
\label{eq:generalizedRD_2d}
\end{equation}
and hypothesized that these two-dimensional systems could also be reformulated with RD systems as follows:

\begin{equation}
    \begin{cases}
      \dfrac{\partial u}{\partial t} = D \nabla^2 u + F(u,\alpha_0 + \sum_{j=1}^M \alpha_j v_j)  \\
      \dfrac{\partial v_j}{\partial t} = \dfrac{1}{\epsilon} (D_j \nabla^2 v_j +\mu u - v_j)
    \end{cases}
    \label{eq:RDemulation}
\end{equation}

This extension is valid if we can find an appropriate set of $\{ \alpha_j \}$ that satisfies the $K*u \simeq \alpha_0 + \sum_{j=1}^M \alpha_j v_j$, although we will not prove mathematically that this is possible for any symmetric 2D kernels.

Having extended it to two-dimensional systems, we can apply this method to the asymptotic Lenia (Eq.\ref{eq:asymLenia_diff}) by setting $D=0$ and $F(u,K*u) = T(K*u) -u $. The shape of non-local kernel $K$ in the asymptotic Lenia is determined by choosing the appropriate $\{\alpha_j\}$.

From these considerations, we constructed our RD system emulation of the asymptotic Lenia (“RD Lenia”) as follows. First, we used $M=40, \mu = 0.1, D_j = j, \epsilon = 0.005$. To approximate $K*u$ by auxiliary variables $\{v_j\}$, we linearly regressed $K*u$ as $K*u \simeq \alpha_0 + \sum_{j=1}^M \alpha_j v_j$ to determine $\{\alpha_j\}$. The size of the time step $dt$ with the Euler method was $dt = 0.00001$. The value of $dt$ and $\epsilon$ have to be very small because, in the case of the original asymptotic Lenia, the computation of the non-local kernel $K*u$ in Eq.\ref{eq:asymLenia} happens at every time step, whereas in this system based on an RD system, we must first wait for $\alpha_0 + \sum_{j=1}^M \alpha_j v_j$ to converge to $K*u$, and this must happen in the time scale smaller than one time step in Eq.\ref{eq:asymLenia}. 

\begin{figure}
\begin{tabular}{cc}
  \begin{minipage}[t]{0.5\hsize}
  \includegraphics[width=\linewidth]{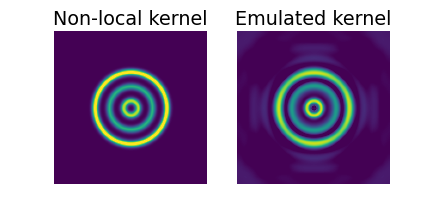}
  \end{minipage} &
 \begin{minipage}[t]{0.4\hsize}
  \includegraphics[width=\linewidth]{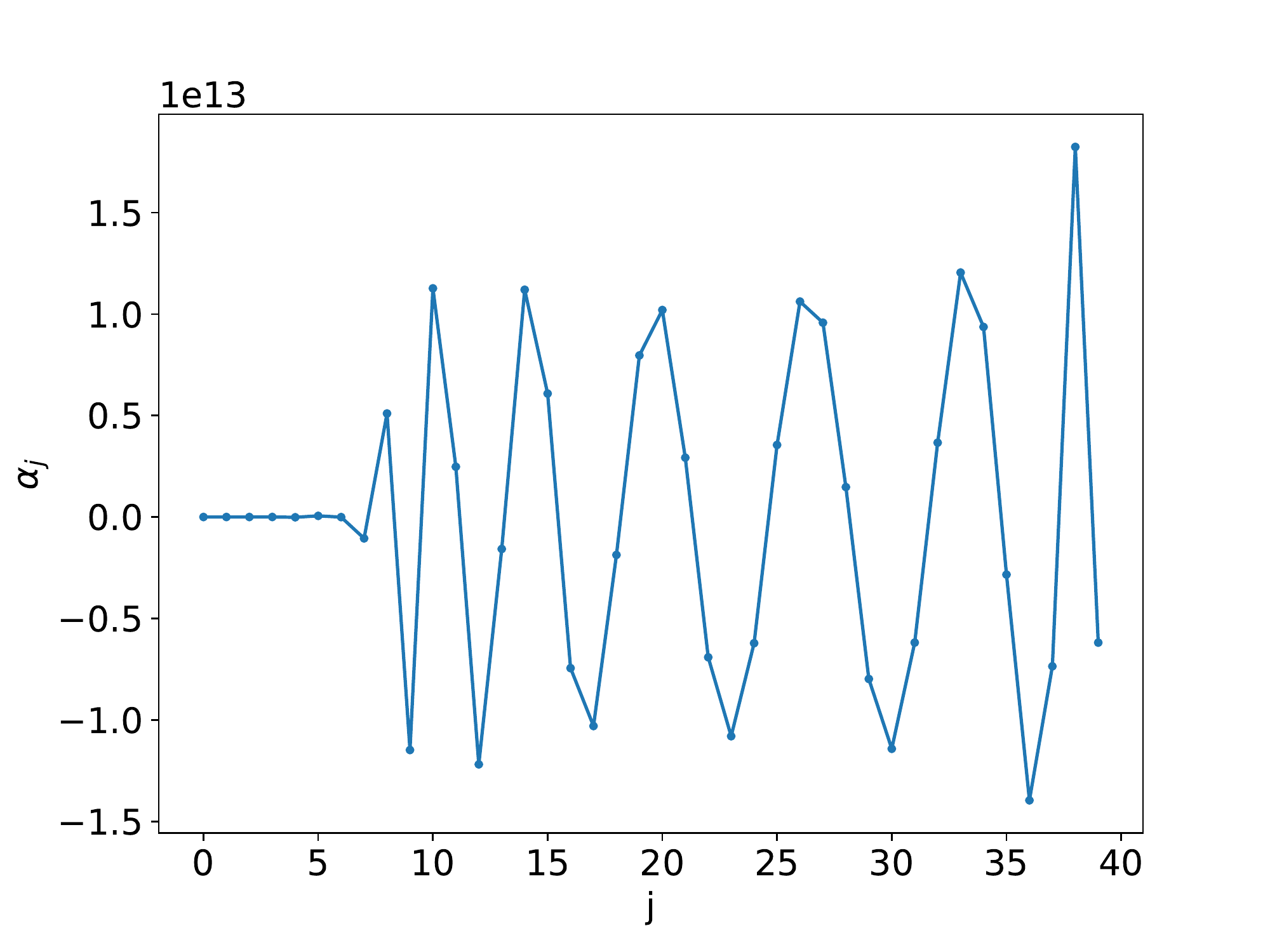}
  \end{minipage} 
\end{tabular}
\caption{Emulation of asymptotic Lenia by RD systems. Left: Original kernel $K(x)$, Middle: Emulated kernel $\alpha_0 + \sum_{j=1}^M \alpha_j v_j(x)$, Right: The values of the  coefficients $\{ \alpha_j \}$ for the emulation of $K(x)$.}
\label{fig:RDemulation_kernel}
\end{figure}

\subsubsection{Implementation of RD Lenia}\label{sec:2.2.2}

First, we checked if this system could emulate the non-local kernel used in the asymptotic Lenia (Fig.\ref{fig:RDemulation_kernel}, Left). To do this, we first fixed $u(x)=\delta(x)$ and ran the system. When we fix $u(x)$, $\{v_j\}$ will converge to certain values, and $\alpha_0 + \sum_{j=1}^M \alpha_j v_j$ should be (nearly) equal to $K*u$, which is equal to $K(x)$ when $u(x)=\delta(x)$. 

After the convergence of $\{v_j\}$, we linearly regressed $K*u$ as $\alpha_0 + \sum_{j=1}^M \alpha_j v_j$ to check how well the linear combination of $\{v_j\}$ reproduced the non-local kernel in the asymptotic Lenia. The original kernel $K$ is shown in Fig.\ref{fig:RDemulation_kernel}(Left) and the emulated kernel derived by linear regression $\alpha_0 + \sum_{j=1}^M \alpha_j v_j$ is shown in Fig.\ref{fig:RDemulation_kernel} (Middle). The coefficient of linear regression, $\{ \alpha_j \}$, is shown in Fig.\ref{fig:RDemulation_kernel} (Right). This confirms that the emulated kernel reproduces the shape of the original non-local kernel $K$ well.

\begin{figure}
\centering
\includegraphics[width=0.4\linewidth]{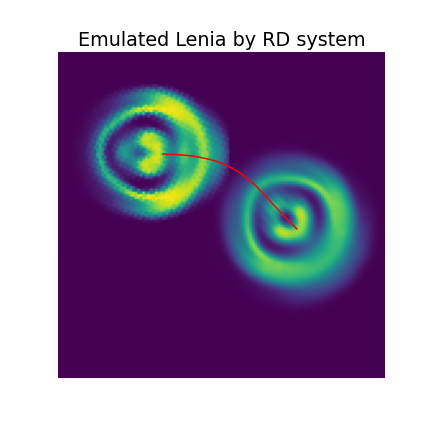}
\caption{Asymptotic Lenia emulated by the RD system, RD Lenia (Eq.\ref{eq:RDemulation}). The initial and final patterns are superimposed. The red line is the motion trajectory calculated as the center of mass of the pattern.}
\label{fig:RDemulation}
\end{figure}

We then simulated the asymptotic Lenia using this emulated RD system. We initialized $u(x,0)$ as the same pattern in the initial pattern of the glider of the asymptotic Lenia and ran the system of Eq.\ref{eq:RDemulation}. The $\{\alpha_j\}$ was also chosen to satisfy $K*u \simeq \alpha_0 + \sum_{j=1}^M \alpha_j v_j$ by linear regression. The result computed over 4,000,000 time steps is shown in Fig.\ref{fig:RDemulation}. The emulation of Eq.\ref{eq:generalizedRD_2d} by Eq.\ref{eq:RDemulation} is only guaranteed if $\epsilon \to 0$ and $K*u = \alpha_0 + \sum_{j=1}^M \alpha_j v_j$. However, in our approximate setting, we have successfully simulated the motion of the glider pattern in the RD system (Fig.\ref{fig:RDemulation}) and confirmed that our RD system is capable of reproducing the behavior of the asymptotic Lenia.

\subsubsection{Can the Reaction Term be implemented by any chemical Reactions?}\label{sec:2.2.3}

We have already shown that the asymptotic Lenia can be formulated as an RD system (RD Lenia), but we also note here that this does not ensure that the system can be implemented by chemical reactions. This is because non-arbitrary reaction terms can be realized by chemical reactions (which typically assume mass action kinetics), and we need to check whether $F(u,\alpha_0 + \sum_{j=1}^M \alpha_j v_j)$ in Eq.\ref{eq:RDemulation} is possible to be implemented as a set of chemical reactions.

The condition to be regarded as mass action kinetics requires that all reaction terms are polynomial and do not contain “negative cross-effects”. Negative cross-effects is a negative term without its own component \citep{hars1981inverse}. For example, the reaction term in chemical species $u$ can include terms like $-u, -u v_1, -u^2 v_1 v_2^2$, where $v_1, v_2$ are other chemical species, but cannot  include terms like $-v_1, -v_1^2, -v_1 v_2^2$ and these terms are called negative cross-effects.

Here, we point out two difficulties in interpreting the reaction term as mass action kinetics in our case. 

The first difficulty relates to the condition that the reaction terms should be polynomial. In the case of asymptotic Lenia, the form of $F$ is $F(u,K*u) = T(K*u) -u = \exp(-(\frac{K*u-m}{s})^2) - u$, which means that we have to approximate the Gaussian function by polynomials. This is in principle possible by Taylor expansion, but it requires a lot of terms to fully approximate the highly localized Gaussian because its convergence is slow. 

The second difficulty relates to the condition of negative cross-effects. Even if we can transform the reaction terms as a linear combination of polynomials, it should not include the negative cross effects. In our case, $F$ is a reaction term for $u$, so this reaction term cannot have a negative term without $u$. However, the Taylor expansion of $F(u,\alpha_0 + \sum_{j=1}^M \alpha_j v_j) = \exp(-(\frac{\alpha_0 + \sum_{j=1}^M \alpha_j v_j - m}{s})^2) - u$ (the values of $\{\alpha_j\}$ are shown in Fig.\ref{fig:RDemulation_kernel}, Right) inevitably contain negative terms without $u$. Therefore, they cannot be interpreted as chemical reactions. 
This difficulty also applies to the original Lenia. As we showed in \S\ref{sec:2.1.3}, the original Lenia does not need the upper clipping. However, we still need the lower clipping and the negative constant term, -1, in the reaction term $G$. The lower clipping does not matter in the chemical system, because negative concentration cannot occur in the chemical system. On the other hand, the negative constant term is problematic because it also corresponds to the negative cross-effects.

These difficulties may be alleviated by introducing generalized chemical kinetics, which includes not only mass-action kinetics but also other kinetics such as Michaelis-Menten kinetics. Nevertheless, it seems difficult to decompose $F$ into chemically plausible terms while satisfying the “strictness condition”, which corresponds to the negative cross-effects condition in generalized settings \citep{fages2015inferring}. 

Because of these difficulties, the RD system we constructed is difficult to be interpreted as a chemical reaction system. However, this does not exclude the possibility that the asymptotic Lenia could be fully described as a chemical RD system using other constructions.

\section{Discussion}
In this paper, we have demonstrated that the asymptotic Lenia formulation is analogous to a generalization of the KT model, a phenomenological model of RD systems. Also, by introducing auxiliary variables, asymptotic Lenia can be formulated as a time-continuous RD system devoid of non-local interactions.

However, the resulting RD system cannot be considered a proper chemical reaction because the reaction terms do not conform to mass-action kinetics. There are two issues to address. The first is the highly localized Gaussian term, which might be overcome by applying generalized chemical reaction kinetics \citep{fages2015inferring} or, fundamentally, by introducing numerous additional reactions into the system. The second problem originates from the kernel approximation, partly accountable for the negative cross-effect. To mitigate this issue, it's imperative to either provide an alternative basis for emulating a non-local kernel without resorting to negative coefficients for approximation, or to emulate the system following a completely innovative approach. Some RD systems, as shown by \citet{ei2021effective}, behave as though they use a non-local kernel, indicating the possibility of a system effectively utilizing a non-local kernel similar to asymptotic Lenia without resorting to negative cross-effects.

Upon overcoming these hurdles, the asymptotic Lenia could be interpreted as chemical RD systems, paving the way for the development of a theory of Lenia grounded on the thermodynamics of RD systems \citep{falasco2018information}  and even permitting the implementation of Lenia in physical space. Conversely, these challenges may be fundamentally insurmountable, suggesting that a chemical implementation of Lenia is unfeasible.

\section{Acknowledgement} We are grateful to Bert Chan for his valuable advice and comments. This project was partly supported by JSPS KAKENHI (21H04885).

\bibliographystyle{unsrtnat}
\bibliography{references}  

\end{document}